\documentclass[preprint,12pt,useAMS]{elsarticle}

\journal{Scientific Reports}

\begin{document}

\begin{frontmatter}



\title{Constraining the axion-photon coupling using radio data of the Bullet Cluster}


\author{Man Ho Chan}

\address{Department of Science and Environmental Studies, The Education University of Hong Kong \\ 
Tai Po, New Territories, Hong Kong, China}

\begin{abstract}
Axion is one of the most popular candidates of the cosmological dark matter. Recent studies considering the misalignment production of axions suggest some benchmark axion mass ranges near $m_a \sim 20$ $\mu$eV. For such axion mass, the spontaneous decay of axions can give photons in radio band frequency $\nu \sim 1-3$ GHz, which can be detected by radio telescopes. In this article, we show that using radio data of galaxy clusters would be excellent to constrain axion dark matter. Specifically, by using radio data of the Bullet cluster (1E 0657-55.8), we find that the upper limit of the axion-photon coupling constant can be constrained to $g_{a \gamma \gamma} \sim 10^{-12}-10^{-11}$ GeV$^{-1}$ for $m_a \sim 20$ $\mu$eV, which is tighter than the limit obtained by the CERN Axion Solar Telescope (CAST).
\end{abstract}

\begin{keyword}
Dark matter
\end{keyword}

\end{frontmatter}



\section{Introduction}
Observational data reveal that some unknown dark matter particles exist in our universe. Some proposals have suggested a class of hypothetical particles called weakly interacting massive particles (WIMPs) which can account for the dark matter \cite{Bertone}. These particles are fermions which may interact with ordinary matter or self-annihilate to give high-energy particles like photons. However, recent direct-detection experiments \cite{Aprile} and large hadron collider experiments \cite{Abecrcrombie} show null result of these particles. Also, a large parameter space of these particles has been ruled out by indirect detections, such as gamma-ray observations \cite{Ackermann,Chan,Abazajian}, cosmic-ray observations \cite{Ambrosi,Aguilar} and radio observations \cite{Egorov,Chan2,Chan3,Chan4}.

On the other hand, some early studies suggested that a class of light scalar or pseudo-scalar particles might exist \cite{Peccei,Weinberg}. In particular, a boson candidate called axion can help solve the CP-violation problem of the strong interaction \cite{Peccei}. Axions are massive and very stable particles which can probably account for the dark matter in our universe \cite{Battye}. Axions can couple with photons which lead to the conversion between axions and photons ($a \rightarrow \gamma$) via the Primakoff effect in the presence of an external electric or magnetic field \cite{Anselm}. They can also decay into photons ($a \rightarrow \gamma+\gamma$), though the decay rate is very low. Examining the coupling between axions and photons provides a possible way for detecting or constraining axions. Some experiments like the International Axion Observatory (IAXO) \cite{Armengaud} and Any Light Particle Search (ALPS) \cite{Bahre} are going to search for the signals of axions. 

Some benchmark cosmological axion models suggest that the axion mass is of the order $m_a \sim 20$ $\mu$eV \cite{Sikivie,Graham,Battye}. For such axion mass, the photons emitted via the spontaneous decay process would be in radio band ($\nu=m_ac^2/2h \sim 2$ GHz). Therefore, using radio data to search for axion is possible. In this article, we show that using radio data of galaxy clusters would be excellent to constrain axion dark matter. We find that the axion-photon coupling constant $g_{a \gamma \gamma}$ can be constrained down to $10^{-12}-10^{-11}$ GeV$^{-1}$ using radio data of the Bullet Cluster (1E 0657-55.8). Our limits are tighter than the CERN Axion Solar Telescope (CAST) limit \cite{Anastassopoulos} by nearly an order of magnitude within $m_a \approx 17-32$ $\mu$eV.

\section{Cosmological decaying axion model}
There are many proposals which suggest axions to account for all cosmological dark matter. The possible mass of axions can range from $10^{-9}$ eV to 10 eV \cite{Battye}. For example, the simplest standard thermal freeze-out arguments suggest that $m_a \approx 4.5-7.7$ eV \cite{Grin,Battye}. Some other models considering symmetry arguments propose $m_a \approx 100-400$ $\mu$eV \cite{Battye2,Battye}. Specifically, some recent studies follow the misalignment mechanism and suggest $m_a \sim 20$ $\mu$eV \cite{Sikivie,Graham}. A more recent calculation of the misalignment production of axions gives \cite{Wantz,Battye}:
\begin{equation}
\Omega_ah_{100}^2 \approx 0.54g_*^{-0.41} \theta_i^2 \left(\frac{m_ac^2}{6~\mu \rm eV} \right)^{-1.19},
\end{equation}
where $g_* \approx 10$ is the number of relativistic degrees of freedom during the realignment process, $\theta_i$ is the initial angle of misalignment, $h_{100}$ is the Hubble parameter defined in $H_0=100h_{100}$ km s$^{-1}$ Mpc$^{-1}$. Following the standard assumption of the mean alignment angle $\theta_a^2=\pi^2/3$ and assuming axions being all cold dark matter ($\Omega_ah_{100}^2 \approx 0.12$), we get $m_a \approx 19-23$ $\mu$eV \cite{Battye}. This proposal with such a narrow range of $m_a$ has become one of the most important benchmark cosmological axion models for further investigation. However, the assumption of the mean alignment angle does not involve the uncertainties from the QCD parameters that connect the axion mass and decay constant. Also, the presence of topological defects might lead to a significantly higher estimate of the dark matter axion mass \cite{Gorghetto}. Therefore, the narrow range of $m_a$ suggested by \cite{Battye} might not be completely justified. Nevertheless, the range of $m_a \sim 10-30$ $\mu$eV being axion dark matter is still a popular range which is worth to have further investigation. In the followings, we will focus on this particular range $m_a \sim 10-30$ $\mu$eV and constrain the relevant parameters.

The frequency of the emitted photons in spontaneous decay of axions ($a \rightarrow \gamma+\gamma$) is given by $\nu=m_ac^2/2h$. Therefore, for $m_a \sim 10-30$ $\mu$eV, the emission frequency is $\nu \sim 1.2-3.6$ GHz. On the other hand, the decay time can be predicted theoretically as \cite{Battye}
\begin{equation}
\tau= \frac{32h}{g_{a \gamma \gamma}^2m_a^3c^6}=8 \times 10^{35}~{\rm s} \left( \frac{g_{a \gamma \gamma}}{10^{-10}~\rm GeV^{-1}} \right)^{-2} \left( \frac{m_ac^3}{250~\mu \rm eV} \right)^{-3}.
\end{equation}
Current general tightest upper limit for the axion-photon coupling constant $g_{a \gamma \gamma}$ is obtained by the CAST solar axion experiment: $g_{a \gamma \gamma}<0.66 \times 10^{-10}$ GeV$^{-1}$ for $m_a<10^{-2}$ eV \cite{Anastassopoulos}. Therefore, for $m_a \sim 10-30$ $\mu$eV, $\tau$ would be larger than $10^{39}$ s, which suggests that axions are very stable compared with the age of our universe ($\sim 4\times 10^{17}$ s). It also reveals that detecting decaying signal of axions might be very difficult. 

Fortunately, the spontaneous decay of axions could be greatly enhanced by the stimulated emission mechanism \cite{Caputo,Caputo2,Battye}. If the background contains a large amount of photons same with the emission frequency, the decay rate would be increased by a large amount. Such enhancement is characterized by the photon occupation number $f_{\gamma}$. The total radio flux density of the spontaneous decay in a structure with total mass $M$ is given by \cite{Battye}
\begin{equation}
S_a=\frac{Mc^2}{4\pi D^2 \tau \Delta \nu}(1+2f_{\gamma}),
\end{equation}
where $D$ is the luminosity distance to the structure, $z$ is the redshift of the structure and $\Delta \nu$ is the frequency width of the decay. There are a few components which can contribute to the photon background to enhance the decay. For example, the cosmic microwave background (CMB) contains a large amount of photons at $\nu \sim 2$ GHz. The photon occupation number for the CMB photons is
\begin{equation}
f_{\gamma}=\left(e^{m_ac^2/2kT_{\rm CMB}}-1 \right)^{-1},
\end{equation}
where $T_{\rm CMB}=2.725$ K. This gives $f_{\gamma} \approx 20-24$. The other components such as synchrotron radio background or hot gas thermal Bremsstrahlung radiation background may be able to give a larger enhancement.

Note that the stimulated decay of axions considered above is a little bit different from the resonant decay of the axion field. As stated in \cite{Blas}, the resonant decay assumes that photons produced from the decays of axions contribute to the photon occupation number responsible for stimulating the decay. Therefore, it will give an exponential growth of photons and the axions would almost completely decay in a short time. For the stimulated decay considered here, the photon occupation number is assumed only to arise from the background photons (e.g. CMB or hot gas). The decaying photons would not further stimulate the decay to trigger exponential growth of photons. Moreover, as discussed in \cite{Arza}, the resonant decay of axions requires that the axion momentum spread is not too large and the gravitational potential well in which axions are bound is not too strong. These constraints might prevent axions from decaying too fast so that most of the axions would remain in the universe \cite{Beck}. However, these conditions do not affect the stimulated decay due to the background photons.

\section{Data analysis}
After an extensive search of archival radio data, we find that the radio data of the Bullet cluster (1E 0657-55.8) reported in \cite{Shimwell} are the best for constraining the benchmark cosmological axion model. It is because it is a large galaxy cluster (the reason for focusing galaxy clusters will be discussed below) and its radio data involve a continuum frequency range covering $\nu \sim 1-3$ GHz with relatively small uncertainties. Some previous studies have also investigated the effect of decaying axion-like dark matter using observational data of the Bullet cluster \cite{Sorensen}. The redshift and the luminosity distance of the Bullet cluster are $z=0.296$ and $D=1529$ Mpc (assuming Hubble parameter $h_{100}=0.7$ and following the standard $\Lambda$CDM model with $\Omega_m=0.3$ and $\Omega_{\Lambda}=0.7$) respectively \cite{Shimwell}. The observing radio frequency is $1.1-3.1$ GHz, including wide band and narrow band (290 MHz sub-bands) methods \cite{Shimwell}. Due to the cosmological redshift, the frequency of the decaying photons would decrease by a factor of $(1+z)$ when the photons arrive the radio telescope, which becomes $\nu_{\rm obs}=\nu(1+z)^{-1}=0.9-2.8$ GHz for $m_a \sim 10-30$ $\mu$eV. Therefore, the observing frequency can cover most of the possible frequency range of the decaying photons.

As mentioned above, the spontaneous decay would be enhanced by the background photons via stimulated emission. In the Bullet cluster, a large amount of hot gas particles exist which are emitting thermal Bremsstrahlung radiation. The amount of radiation can be characterized by the hot gas temperature $T$. For thermal Bremsstrahlung emission, the hot gas is in local thermodynamic equilibrium such that the emissivity $j_{\nu}$ is the product of the absorptivity $k_{\nu}$ and the radiation spectral intensity $I_{\nu}$ (i.e. $j_{\nu}=k_{\nu}I_{\nu}$). In view of this, the hot gas particles and photons are interacting via absorption and emission processes. The radiation spectral intensity $I_{\nu}$ represents the background photon distribution in the hot gas and it follows the Planck spectrum: $I_{\nu}=(2h\nu^3/c^2)[\exp(h\nu/kT)-1]^{-1}$. Note that the Bremsstrahlung emission (the photons escaped from the hot gas to reach us) does not follow the Planck spectrum and it mainly consists of X-ray photons (represented by the emissivity $j_{\nu}$) because the optical depth is very small for X-ray photons. However, our focus is the photons inside the hot gas, not the photons escaped from the hot gas. The background photon distribution inside the hot gas volume still maintains the Planck spectrum via absorption and emission equilibrium (represented by the spectral intensity $I_{\nu}$). Therefore, the photon occupation number in the hot gas is also 
\begin{equation}
f_{\gamma}=\left(e^{m_ac^2/2kT}-1 \right)^{-1} \approx \frac{2kT}{m_ac^2}.
\end{equation}
For the Bullet cluster, we have $T=17.4 \pm 2.5$ keV \cite{Tucker,Hayashi} and $f_{\gamma} \sim 10^9$. Therefore, the stimulated emission can be greatly enhanced in the thermal Bremsstrahlung hot gas. For the synchrotron radio background, the corresponding photon occupation number is just $f_{\gamma} \sim 0.1$, which is not significant. A simple physical intuition may think that the peak energy of thermal Bremsstrahlung emission is of the order keV, which is much higher than the photon energy required for stimulated emission ($\sim 10^{-5}$ eV) so that the hot gas component may have less effect on stimulated emission. However, the background photon distribution of the hot gas (not the escaped X-ray photons) also follows the Planck spectral distribution, which gives a very large photon occupation number at low energy. This is because the absorption of low-energy photons ($\sim 10^{-5}$ eV) in the keV-temperature hot gas is still important. For the synchrotron radio background component, the power-law distribution suppresses the photon occupation number so that this component does not give a significant enhancement in stimulated emission.

The frequency width $\Delta \nu$ can be written in terms of the velocity dispersion of axion dark matter $\sigma_a$: $\Delta \nu=\nu_{\rm obs}(\sigma_a/c)$ \cite{Caputo}. The average velocity dispersion of axion dark matter can be found by the Virial theorem:
\begin{equation}
\sigma_a= \sqrt{\frac{3kT}{\mu m_p}},
\end{equation}
where $\mu=0.59$ is the molecular weight and $m_p$ is the proton mass. Putting $T=17$ keV, we get $\sigma_a \approx 2800$ km/s, which gives a very narrow frequency width $\Delta \nu \sim 0.02$ GHz. Therefore, a very sharp radio line would be observed if the spontaneous decay is strong enough. Due to the narrow frequency width, the radio data observed or considered must be in wide band or continuous band. Otherwise, the possible radio signal of the decay would be ignored if the observing frequencies do not match the specific decaying photon frequency.

From the whole radio spectrum ($\nu_{\rm obs}=1.1-3.1$ GHz) reported in \cite{Shimwell}, we can analyse the axion decay for $m_a \approx 14-32$ $\mu$eV. This range of $m_a$ would be divided into 6 sub-bands because there are 6 sub-bands of frequencies in the radio observations. For the narrow mass range $m_a=19-23$ $\mu$eV suggested in \cite{Battye}, there are two continuous sub-bands (1.6-1.9 GHz and 1.9-2.2 GHz) which can cover the observing frequency range of the axion decay (1.8-2.1 GHz). 

Firstly, we can get the most conservative limit of the axion-photon coupling constant if we assume that all radio flux density detected in these sub-bands originates from axion decay. Putting $z=0.296$, $D=1529$ Mpc \cite{Shimwell} and the virial mass of the Bullet cluster $M=3.1 \times 10^{15}M_{\odot}$ \cite{Hayashi} into Eq.~(3), we get
\begin{equation}
S_a=66.5~{\rm mJy} \left(\frac{g_{a \gamma \gamma}}{10^{-10}~\rm GeV^{-1}} \right)^2 \left(\frac{m_ac^2}{20~\mu \rm eV} \right)^3 \left(\frac{\nu_{\rm obs}}{2~\rm GHz} \right)^{-1} \left(\frac{1+2f_{\gamma}}{10^9} \right).
\end{equation}
Note that the value of the luminosity distance $D$ is calculated based on the standard $\Lambda$CDM model. The uncertainty of $D$ would not be very significant because the uncertainty of the observed redshift $z$ is small. However, the uncertainty in $M$ and $T$ might have some impact on the calculation of $S_a$ as $S_a$ is directly proportional to $M$ and $T^{1/2}$. Here, the percentage uncertainty in $T$ is about 14\% \cite{Tucker}, which would only give about 7\% uncertainty in $S_a$. Nevertheless, the value of $M$ is somewhat model-dependent, which might give a significant systematic uncertainty in $S_a$. The value of $M$ adopted in this study is calculated based on the parameters following the Navarro-Frenk-White (NFW) dark matter density profile \cite{Clowe,Hayashi}, which is one of the most robust profiles to describe the dark matter density profiles in galaxy clusters \cite{Pointecouteau}.
 
The $1\sigma$ radio flux density upper limits for the frequency range $\nu_{\rm obs}=1.1-3.1$ GHz have been obtained from observations \cite{Shimwell}. We can therefore obtain the conservative upper limit of $g_{a \gamma \gamma}$ for different $m_a$ using Eq.~(7) (see Fig.~1). The conservative upper limit is $g_{a \gamma \gamma} \approx (2.3-5.3)\times 10^{-11}$ GeV$^{-1}$, which is tighter than the limit obtained by CAST ($g_{a \gamma \gamma}<6.6 \times 10^{-11}$ GeV$^{-1}$) \cite{Anastassopoulos}. 

Nevertheless, the synchrotron radiation of cosmic rays in a galaxy cluster usually dominates the radio emission. Thus, assuming that all radio flux density originates from axion decay is unrealistic. Moreover, considering the cosmic-ray contribution can further constrain the upper limit of $g_{a \gamma \gamma}$. Using the radio flux density of other sub-bands can predict the background radio emission due to cosmic rays. Based on the radio spectrum of the Bullet cluster obtained in \cite{Shimwell}, the entire spectrum between 1.1-3.1 GHz can be best fitted by a power-law spectrum $S_{\rm CR}=S_0 \nu_{\rm obs}^{-\alpha}$ with a constant spectral index $\alpha=1.43 \pm 0.15$. This would become our null hypothesis for comparison (no axion decay scenario).

We now examine a two-component model: cosmic-ray contribution plus axion decay contribution. The total observed radio flux density should be a sum of two components $S_{\rm tot}=S_a+S_{\rm CR}$. Here, we assume that $S_a$ is a sharp Gaussian function centred at the central frequency of each sub-band with frequency width $\Delta \nu$. Here, we consider that the axion decay component is non-zero for $\nu_{\rm obs}=1.1-3.1$ GHz only. Moreover, the spectral index $\alpha$ and the normalization constant $S_0$ are set as free parameters because adding the axion decay component would slightly alter these values obtained from the null hypothesis. Therefore, there are three free parameters ($S_0$, $\alpha$ and $g_{a \gamma \gamma}$) in the spectral fits. The goodness of fits can be examined by the $\chi^2$ value defined as
\begin{equation}
\chi^2= \sum_i \frac{(S_{\rm tot,i}-S_i)^2}{\sigma_i^2},
\end{equation}
where $S_i$ and $\sigma_i$ are the observed radio flux density and their uncertainties respectively. As mentioned in \cite{Dafni}, the Chernoff's theorem states that the test statistic is asymptotically distributed according to $0.5\chi^2+0.5\delta(0)$ when the null hypothesis is true \cite{Chernoff,Cowan}. Therefore, based on the test statistic value, we can determine the corresponding statistical significance. 

After considering the two-component model, we can get the best-fit scenarios for different sub-bands $m_a$ (see Table 1). In particular, we find that there exist strong possible signatures of axion decay in the sub-bands $m_a=14-17$ $\mu$eV and $m_a=29-32$ $\mu$eV. The statistical significance is $4.2\sigma$ and $3.4\sigma$ respectively compared with the null hypothesis. The overall best-fit axion-photon coupling constant is $g_{a \gamma \gamma}=2.4 \times 10^{-11}$ GeV$^{-1}$ at $m_a \approx 15.3$ $\mu$eV (with $\chi^2=16.4$). Furthermore, we can also release $m_a$ as a free parameter to get a $2\sigma$ contour of ($m_a$, $g_{a \gamma \gamma}$) (i.e. with $\chi^2 \le 22.6$ for 2 degrees of freedom). The region inside the contour indicate the $2\sigma$ range of ($m_a$, $g_{a \gamma \gamma}$) from the best-fit parameters (see Fig.~1). Note that we have neglected the look-elsewhere effect here as we have constrained our range of $m_a$ in the analysis. We also show the spectral fits of the two-component model in Fig.~2 for the best-fit parameters. The peaks indicate the contribution of axion decay for the best-fit scenarios for different sub-bands. Besides, we can also determine the upper limits of $g_{a \gamma \gamma}$ ruled out at $2\sigma$ by comparing the two-component model with the null hypothesis (no axion decay). We assume the central value of $m_a$ for each mass bin (i.e. frequency bin) to get the upper limits of $g_{a \gamma \gamma}$ and corresponding parameters (see Table 2). By including the cosmic-ray emission, the upper limits of $g_{a \gamma \gamma}$ can be further constrained down to $\sim 10^{-12}$ GeV$^{-1}$ (see Fig.~1). 

\begin{figure}
\vskip 5mm
 \includegraphics[width=140mm]{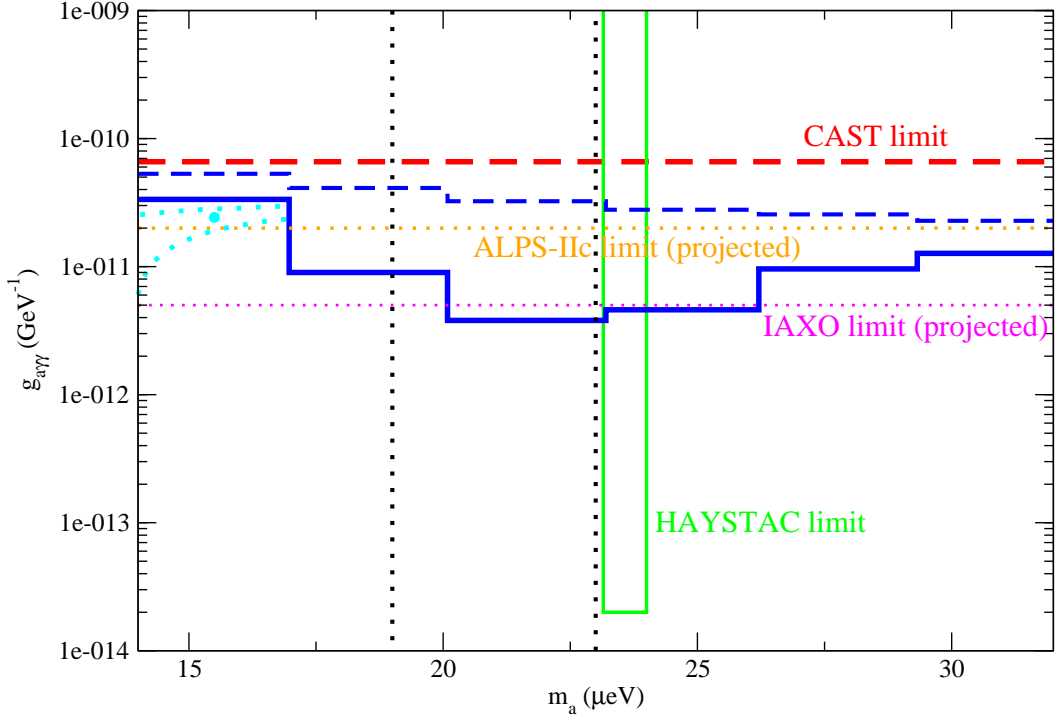}
\caption{The blue dashed line and blue solid line represent the conservative upper limit of $g_{a \gamma \gamma}$ and the two-component model upper limit of $g_{a \gamma \gamma}$ (with the mean value of $m_a$ in each bin) respectively. The cyan dot represents the overall best-fit $g_{a \gamma \gamma}$ across all $m_a$ bins. The region inside the cyan dotted contour is the $2\sigma$ range from the overall best-fit parameters. The region between the black dotted lines indicate the narrow mass range $m_a=19-23$ $\mu$eV suggested in \cite{Battye}. The red dashed line, orange dotted line and the pink dotted line indicate the upper limits form the CAST experiment \cite{Anastassopoulos}, projected upper limit of ALPS-IIc \cite{Bahre} and projected upper limit of IAXO \cite{Graham} respectively. The region bounded by the green lines represent the HAYSTAC ruled out region \cite{Zhong}.}
\label{Fig1}
\vskip 3mm
\end{figure}

\begin{figure}
\vskip 5mm
 \includegraphics[width=140mm]{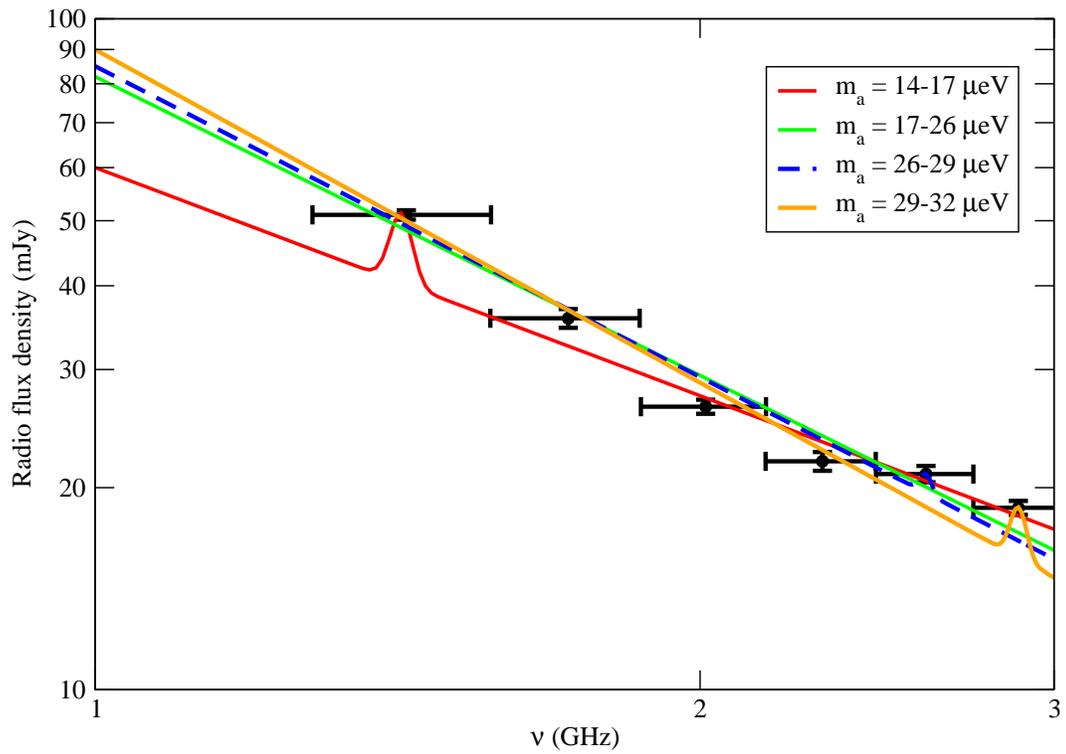}
\caption{The data points with error bars of the Bullet cluster radio spectrum are extracted from \cite{Shimwell}. The coloured lines show the spectra of the two-component model with the best-fit parameters for different sub-bands of $m_a$.}
\label{Fig2}
\vskip 3mm
\end{figure}

\section{Discussion}
In this article, we have used the radio spectral data of the Bullet cluster to investigate the potential axion decay signal and constrain the axion-photon coupling constant. We have found the overall best-fit scenario (the largest statistical significance compared with the null hypothesis) which is for the sub-band $m_a=14-17$ $\mu$eV. The radio excess in this sub-band may indicate a potential signal of axion decay. Besides, we further constrain the axion-photon coupling constant $g_{a \gamma \gamma}$ down to $\sim 10^{-12}-10^{-11}$ GeV$^{-1}$ for the popular axion mass range $m_a\sim 10-30$ $\mu$eV. The upper limits obtained are tighter than the CAST limit, especially for the mass range $m_a=19-23$ $\mu$eV suggested in \cite{Battye}.

In fact, some experiments are going to search for the signal of axions. For example, the IAXO and ALPS are performing experiments to search for axions and constrain the parameters of axions. The projected limits of $g_{a \gamma \gamma}$ constrained by these experiments are $g_{a \gamma \gamma}<5 \times 10^{-12}$ GeV$^{-1}$ (IAXO) \cite{Graham} and $g_{a \gamma \gamma}<2 \times 10^{-11}$ GeV$^{-1}$ (ALPS-IIc) \cite{Bahre} respectively. Our upper limits for $m_a=14-32$ $\mu$eV are somewhat tighter than the ALPS-IIc projected limit and close to the IAXO limit (see Fig.~1). Therefore, using radio data of galaxy clusters is excellent to constrain axion decay as the limits can be comparable to the current axion searching experiments. It can be used as a complementary measurement for the axion search.

Some specific models suggest $g_{a \gamma \gamma} \sim 10^{-14}$ GeV$^{-1}$ for $m_a\sim 20$ $\mu$eV \cite{diCortona}. Nevertheless, our upper limits obtained are still far from this predicted order of magnitude. Although some detections such as the ADMX haloscope \cite{Asztalos,Hoskins} and HAYSTAC microwave cavity axion experiment \cite{Zhong} can constrain it down to $g_{a \gamma \gamma} \sim 10^{-15}-10^{-14}$ GeV$^{-1}$. However, the axion mass ranges constrained for these detections are $m_a=1.90-3.69$ $\mu$eV (ADMX) and $m_a=23.1-24.0$ $\mu$eV (HAYSTAC) respectively, which only cover a certain narrow mass ranges. Future plan of HAYSTAC experiment might be able to cover the entire range ($m_a=0.5-40$ $\mu$eV) \cite{Simanovskaia}. Besides, based on this current study, future radio observations of galaxy clusters with a very good sensitivity at frequency $\nu_{\rm obs} \sim 2$ GHz (e.g. using the Square Kilometer Array) could also be helpful in detecting axion decay signals or constraining axion-photon coupling constant down to $10^{-14}$ GeV$^{-1}$.

\begin{table}
\caption{The fitting parameters of the null hypothesis ($M_1$ model) and the best-fit two-component model ($M_2$ model). Here, the value of $g_{a \gamma \gamma}$ is the best-fit value for each $m_a$ sub-band.}
 \label{table1}
 \begin{tabular}{@{}lccccc}
  \hline
  Model & $m_a$ & $S_0$ & $\alpha$ & $g_{a \gamma \gamma}$ & $\chi^2$ \\
		& ($\mu$eV) & (mJy) &		& (GeV$^{-1}$) &   \\
  \hline
  \hline
  $M_1$ &  & 82 & 1.48 &  & 50.9 \\
  \hline
  $M_2$ & 14-17 & 60 & 1.13 & $2.4 \times 10^{-11}$ & 16.4 \\
	    & 17-20	& 82 & 1.48 & 0 & 50.9 \\
	    & 20-23 & 82 & 1.48 & 0 & 50.9 \\
	    & 23-26 & 82 & 1.48 & 0 & 50.9 \\
	    & 26-29 & 85 & 1.54 & $6.3 \times 10^{-12}$ & 48.7 \\
	    & 29-32 & 90 & 1.65 & $8.9 \times 10^{-12}$ & 28.8 \\
  \hline
 \end{tabular}
\end{table}

\begin{table}
\caption{The fitting parameters of the two-component model ($M_2$ model) just ruled out at $2\sigma$ compared with the null hypothesis (with the mean value of $m_a$ in each bin). Here, the values of $g_{a \gamma \gamma}$ are the upper limits.}
 \label{table2}
 \begin{tabular}{@{}lccc}
  \hline
  $m_a$ & $S_0$ & $\alpha$ & $g_{a \gamma \gamma}$ \\
  ($\mu$eV) & (mJy) &		& (GeV$^{-1}$) \\
  \hline
  14-17 & 44 & 0.79 & $3.4 \times 10^{-11}$ \\
  17-20	& 83 & 1.49 & $9.0 \times 10^{-12}$ \\
  20-23 & 82 & 1.48 & $3.8 \times 10^{-12}$ \\
  23-26 & 84 & 1.51 & $4.6 \times 10^{-12}$ \\
  26-29 & 88 & 1.62 & $9.6 \times 10^{-12}$ \\
  29-32 & 96 & 1.77 & $1.3 \times 10^{-11}$ \\
  \hline
 \end{tabular}
\end{table}

\section{Acknowledgements}
We thank the anonymous referees for useful constructive feedbacks and comments. The work described in this paper was partially supported by the Seed Funding Grant (RG 68/2020-2021R) and the Dean's Research Fund of the Faculty of Liberal Arts and Social Sciences, The Education University of Hong Kong, Hong Kong Special Administrative Region, China (Project No.: FLASS/DRF 04628).





\end{document}